# Acoustic Energy Storage in Single Bubble Sonoluminescence

Michael P. Brenner,[1] Sascha Hilgenfeldt,[2] Detlef Lohse,[2] and Rodolfo R. Rosales[1]

[1]*Department of Mathematics, Massachusetts Institute of Technology, Cambridge, Massachusetts 02139*
[2]*Fachbereich Physik der Universität Marburg, Renthof 6, 35032 Marburg, Germany*
(Received 19 April 1996)

Single bubble sonoluminescence is understood in terms of a shock focusing towards the bubble center. We present a mechanism for significantly enhancing the effect of shock focusing, arising from the storage of energy in the acoustic modes of the gas. The modes with strongest coupling are not spherically symmetric. The storage of acoustic energy gives a framework for understanding how light intensities depend so strongly on ambient gases and liquids and suggests that the light intensities of successive flashes are highly correlated.    [S0031-9007(96)01412-3]

Sonoluminescence (SL), the conversion of acoustic energy into light, occurs by coupling gaseous bubbles to an externally forced liquid [1]. Two experimental configurations for SL exist: multibubble sonoluminescence (MBSL) [1,2], which occurs in transient cavitation clouds, and single bubble sonoluminescence (SBSL), which occurs when a single bubble is trapped at the node of an applied acoustic field [3,4]. Recent experiments [3–10] uncovered many remarkable properties of SBSL, including picosecond light pulses [4,11] and sensitive dependences on almost all experimental parameters.

SL requires both energy transfer from the liquid to the gas, and focusing of this energy. The shock theory stipulates that a shock focuses the energy input during a single collapse. Large temperatures occur because a focusing shock is a *singular* solution to the Euler equations [12] in which the maximum temperature diverges at the focusing point. Although this shock theory gives consistent explanations for many aspects of SL (e.g., high temperatures and picosecond light pulses), there is mounting experimental evidence that tensions in the theory exist. To wit: While the shock theory is believed to work for both MBSL [13] and SBSL [14], the entire MBSL bubble cloud emits less than 1% of the light of a single bubble in SBSL. For SBSL in alcohols, Weninger *et al.* [9] reported the existence of *abrupt* jumps in the light intensity, in which the light output changes by a factor of 400 with a 1 °C change in the liquid temperature. More recently, Weninger *et al.* [10] reported the existence of angular correlations in the light output, in which the radiation field had a significant dipole moment. We believe that it is difficult to resolve these discrepancies within the single shock theory, and that additional physics is needed.

This paper presents a mechanism for enhancing the effect of shock focusing, giving natural resolutions to the aforementioned tensions in the shock theory. The idea is that the energy focused in SBSL is not input in a single bubble collapse, but instead accumulates within the acoustic modes of the gas over many bubble cycles. Because the energy stored in the modes is far greater than the energy input in a single collapse, the focusing power of the shocks is significantly enhanced. (Note that the emitted light carries away very little energy, and merely acts as an indicator of the energy density at collapse [14].) Within this picture, the maximum temperature and light intensity of SBSL are largely set by the total acoustic energy stored within the bubble, instead of the details of the focusing singularity. An essential condition for accumulative energy storage is the stability of the bubble [4] from surface, diffusive, and chemical instabilities [6,15,16], which exists in SBSL but not in MBSL.

Our theory is motivated by the fact that the basic assumption of the shock theory—that the shock launches into a gas with a spatially uniform density—breaks down in the absence of dissipation. Without dissipation, energy transferred from the liquid to the gas in a single collapse never leaves the bubble, resulting in a nonuniform gas density over long times. This inconsistency is cured only with enough dissipation so that the gas dynamics is *overdamped*. When the system is *underdamped* energy builds up in the acoustic modes of the bubble.

The key quantity distinguishing the underdamped and overdamped regimes is the Floquet multiplier $\Lambda$, characterizing the net input of energy to the acoustic modes in a single cycle. To be self-consistent, the shock theory requires $\Lambda < 1$; acoustic energy storage corresponds to $\Lambda > 1$. The total multiplier is the product $\Lambda = \lambda D$ of multipliers for energy input $\lambda$ and dissipation $D < 1$. Energy input occurs mainly at the bubble collapse through resonances and shocks. Dissipation occurs throughout the cycle and is the product of several factors, mainly viscous dissipation in the gas and acoustic energy transmission from the bubble to the fluid.

In the following, we first compute the acoustic modes of the gas, and then address $\Lambda$ for the different modes. It emerges that the two most easily excited modes are not spherically symmetric. Then, we discuss dissipation. We argue that varying the dissipation naturally leads to *transitions* between the underdamped $\Lambda > 1$ and overdamped $\Lambda < 1$ regimes, manifest by an abrupt jump in the light intensity, as seen by Weninger [9]. Finally, we present further comparisons and predictions for experiments.

We now turn to a calculation of the normal modes of the gas. To keep it simple, assume nondissipative, isothermal gas dynamics in the bubble with a velocity $\boldsymbol{v}(\boldsymbol{r},t)$ and gas density $\rho(\boldsymbol{r},t)$,

$$\partial_t \rho = -\nabla \cdot (\rho \boldsymbol{v}), \qquad \partial_t \boldsymbol{v} + \boldsymbol{v} \cdot \nabla \boldsymbol{v} = -c_{\text{gas}}^2 \frac{\nabla \rho}{\rho}. \quad (1)$$

Here, $c_{\text{gas}}$ is the (constant) sound speed which is determined by the gas temperature $\Theta_g$. The boundary condition for the velocity is set by the Rayleigh-Plesset equation [6,17]

$$R\ddot{R} + \frac{3}{2}\dot{R}^2 = \frac{1}{\rho_l}[p(R,t) - P(t) - P_0] + \frac{R}{\rho_l c_l}\frac{d}{dt} \\ \times [p(R,t) - P(t)] - 4\nu\frac{\dot{R}}{R} - \frac{2\sigma}{\rho_l R}, \quad (2)$$

which describes the dynamics of the bubble wall radius $R(t)$, with $\boldsymbol{v}(R,t)\cdot\hat{\boldsymbol{r}} = \dot{R}(t)$. Typical parameters for an argon bubble in water [8] are surface tension $\sigma = 0.073$ kg/s$^2$, water viscosity $\nu = 10^{-6}$ m$^2$/s, density $\rho_l = 1000$ kg/m$^3$, and speed of sound $c_l = 1481$ m/s. The driving frequency of the acoustic field $P(t) = P_a \cos \omega t$ is $\omega/2\pi = 26.5$ kHz [8] and the external pressure $P_0 = 1$ atm. The pressure inside the bubble varies as $p(R) \propto (R^3 - h^3)^{-\kappa}$, with $h$ the hard core van der Waals radius. Here $\kappa$ is the effective polytropic constant, determined by thermal effects to be $\kappa \approx 1$ [18].

The density and velocity consistent with the Rayleigh-Plesset equation are

$$\rho_s = \rho_0\left(\frac{R_0}{R}\right)^3 + O(M^2), \qquad \boldsymbol{v}_s = \frac{\dot{R}}{R}r\hat{\boldsymbol{r}} + O(M^2), \quad (3)$$

where $M$ is the Mach number $M = \dot{R}/c_{\text{gas}}$. The normal modes follow from linearizing about this state by writing $\rho = \rho_s + \tilde{\rho}(\boldsymbol{r}/R,t)$ and $\boldsymbol{v} = \boldsymbol{v}_s + \tilde{\boldsymbol{v}}(\boldsymbol{r}/R,t)$. Then

$$\partial_t(\tilde{\rho}/\rho_s) = -\nabla_x \cdot \tilde{\boldsymbol{v}}/R, \quad (4)$$

$$\partial_t \tilde{\boldsymbol{v}} + \frac{\dot{R}}{R}\hat{\boldsymbol{r}} \cdot \tilde{\boldsymbol{v}}\hat{\boldsymbol{r}} = -\frac{c_{\text{gas}}^2}{R}\nabla_x\left(\frac{\tilde{\rho}}{\rho_s}\right), \quad (5)$$

where $\boldsymbol{x} = \boldsymbol{r}/R$, and $\nabla_x$ is the gradient with respect to $\boldsymbol{x}$. Introducing the velocity potential via $\tilde{\boldsymbol{v}}(\boldsymbol{x},t) = \nabla\Phi(\boldsymbol{r},t)$ yields a wave equation for $\Phi(\boldsymbol{x},t)$,

$$\partial_t^2 \Phi = c_{\text{gas}}^2 \nabla_x^2 \Phi / R^2. \quad (6)$$

The general solution to this equation is a sum of modes $\Phi(\boldsymbol{r},t) = \sum_{l,n} a_{l,n}(t) j_l(k_{l,n} r/R) Y_l$, where $j_l$ are spherical Bessel functions, $Y_l$ are spherical harmonics depending only on the angle variables, and the $k_{l,n}$ are determined by the boundary condition for the radial velocity $\tilde{\boldsymbol{v}}(R,t)\hat{\boldsymbol{r}} = 0$, i.e., the $n$th zeros $j_l'(k_{l,n}) = 0$. The coefficients $a_{l,n}$ satisfy the equation for a linear oscillator with time varying frequency (often called Hill equation)

$$\ddot{a}_{l,n} + \Omega_{l,n}^2 a_{l,n} = 0, \quad (7)$$

with $\Omega_{l,n} = c_{\text{gas}} k_{l,n}/R$.

When there is a large separation of scales between the oscillator frequency $\Omega_{l,n}$ and $\dot{R}/R$, no energy is fed into $a_{l,n}$; the oscillator energy divided by its frequency, and therefore the envelope $I$ of $a_{l,n}^2/R$ is an adiabatic invariant [19]. When the separation of scales is violated ($\dot{R} \sim c_{\text{gas}} k_{l,n}$), energy is transferred into $a_{l,n}$ and $I$ is no longer invariant, but increases by a factor $\lambda$. The most efficient energy transfer (largest $\lambda$) happens for the modes with the *smallest* values of $k_{l,n}$. The first three are $k_{1,1} \approx 2.0816$, $k_{2,1} \approx 3.3421$, and $k_{0,1} \approx 4.4934$. Thus, the two most easily excitable modes are not spherically symmetric [20].

To demonstrate the energy buildup in acoustic modes, Fig. 1 shows the most unstable mode $a_{1,1}(t)$, together with $R(t)$ and $a_{1,1}^2/R$, clearly demonstrating the parametric instability of Eq. (7). The fast time scale in $a_{1,1}(t)$ corresponds to the acoustic travel time.

This energy buildup eventually saturates due to dissipation and nonlinear effects. We do not understand the detailed structure of the saturated state. However, in general, many different modes $a_{l,n}$ will be excited and interact. The shock produced by the bubble collapse will focus the total energy contained in this saturated state. This shock is much stronger than when only the energy input during a single collapse is focused; thus, the light intensity is significantly enhanced. Since shock focusing in spherical geometry is generically singular, we still expect that there are qualitative similarities between shock focusing in this highly nonuniform gas and shock focusing in a homogeneous gas [12,14]. In particular, the width of the light pulse within this picture should be comparable to the width in the shock picture.

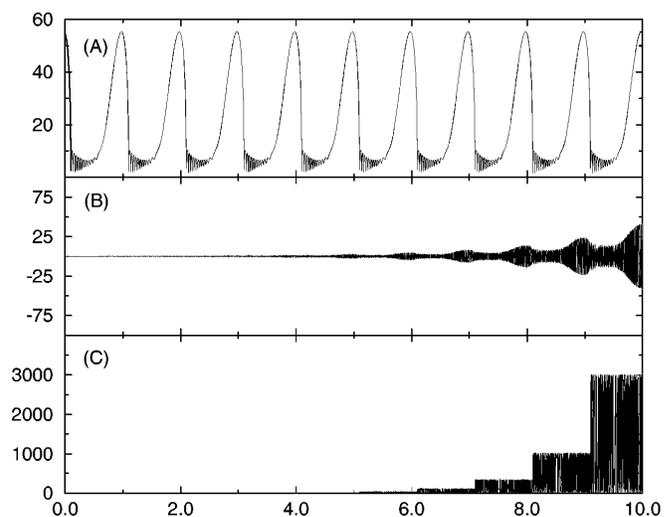

FIG. 1. (a) Solution $R(t)$ to the Rayleigh-Plesset equation. The forcing pressure is $P_a = 1.4$ atm and the ambient radius $R_0 = 7$ $\mu$m. (b) Buildup of energy in the normal mode $a_{1,1}(t)$. (c) $a_{1,1}^2/R$ whose envelope is the adiabatic invariant $I$.

We now proceed to analyze parameter dependences of this acoustic resonator mechanism. The above model neglects dissipation. However, in the real system two major types of dissipation are present, namely, viscous dissipation and acoustic wave transmission from the gas into the liquid. First we consider viscous dissipation: a viscosity $\mu$ causes the adiabatic invariant to decrease between successive bubble collapses by a factor $D_v$. For a mode of wave number $k$, $\dot{I} \sim -\mu/\rho (k/R)^2 I$, so that

$$D_v \sim \exp\left[-\int_0^T \frac{\mu}{\rho_s}\frac{k^2}{R^2}\,dt\right], \tag{8}$$

where $T$ is the bubble oscillation period. Since the bubble spends most of its time near the maximum radius $R_{\max}$, we estimate $D_v \sim \exp[-Tk^2\mu/R_{\max}^2\rho(R_{\max})]$. A crucial point is that $\rho(R_{\max})$ is not determined by Eq. (3) for $\rho_s$ but instead by the *vapor pressure* $p_v$ of the liquid. This parameter depends on the liquid temperature $\Theta_l$ via $p_v \propto e^{-E_a/k_B\Theta_l}$ (with $E_a$ an activation energy).

Another dependence of the Floquet multiplier $\Lambda$ on $\Theta_l$ comes through $\lambda$. The energy input $\lambda$ becomes larger for smaller $c_{\text{gas}}$, since decreasing $c_{\text{gas}}$ increases the time that the liquid and gas frequencies are comparable. Numerical simulations of Eq. (7) show that $\lambda$ decays exponentially with $c_{\text{gas}}$. Since the gas temperature before the bubble collapse is equal to the liquid temperature [6,18], we have $c_{\text{gas}} \propto \sqrt{\Theta_g} = \sqrt{\Theta_l}$. Combining these results, we can now plot the $\Theta_l$ dependence of the total Floquet multiplier $\Lambda(\Theta_l) = \lambda(\Theta_l)D_v(\Theta_l)$, which is shown in Fig. 2(a) [21]. When $\Theta_l$ is decreased, $\Lambda$ increases up to some temperature where the light intensity is maximal. At a critical temperature $\Theta_l^*$ [indicated by the arrow in Fig. 2(a)] dissipation beats the input and a transition towards the overdamped regime occurs, manifested by an abrupt change in the acoustic energy stored by the bubble, and thus in the light intensity.

As mentioned above, this kind of behavior is found by Weninger et al. [9] in experiments on alcohols [Fig. 2(b)]. Decreasing $\Theta_l$, the light intensity increases to a maximum, followed by an abrupt transition. In this transition, the light intensity decreases by a factor of 400 with a 1 °C change in $\Theta_l$. The above argument suggests that the crossover temperature $\Theta_l^*$ should be largest for the fluid with the smallest vapor pressure, all other parameters being constant. This is confirmed by the experiments [9]: of the five alcohols tried, transitions occur in 1-pentanol and 1-butanol, with the lowest vapor pressures at room temperature (0.003 and 0.007 atm, respectively) [22].

Slightly above the temperature $\Theta_l^*$ of the abrupt transition, the Floquet multiplier is $\Lambda \approx 1 + \alpha(\Theta_l - \Theta_l^*)$. The number of cycles $N$ it takes for the system to saturate satisfies $\Lambda^N \propto E_{\text{sat}}$, with $E_{\text{sat}}$ the saturation energy. This gives that $N$ diverges like $N \propto (\Theta_l - \Theta_l^*)^{-1}$. The theory also suggests that below the sharp transition the SBSL intensity should be comparable to that in MBSL, since single shocks are responsible for the light intensity.

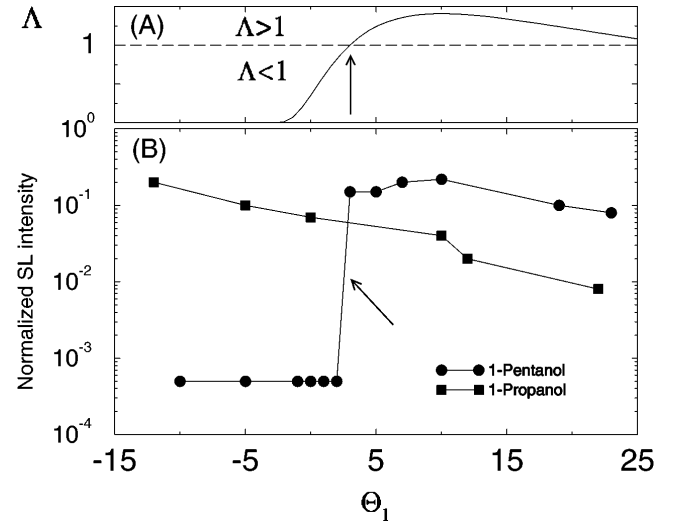

FIG. 2. (a) A sketch of the qualitative behavior of the Floquet multiplier $\Lambda(\Theta_l)$. Scales are adopted to match (b): Dependence of SL intensity on $\Theta_l$ for a xenon gas bubble at a partial pressure of 150 mm Hg for two alcohols, as measured by [9]. The transition turning off the collective acoustic state is marked on the graph. 1-propanol shows no transition in this temperature regime because its vapor pressure is larger than that of 1-pentanol. Similar transitions have also been observed in butanol at slightly different parameters.

We also speculate that, due to the smaller central temperatures of bubbles below the transition, SBSL *spectra* may show more structure in this regime, perhaps resembling those of MBSL.

Whether or not the abrupt transition actually occurs in a given system depends on (a) the value of $\Lambda$ at the liquid freezing-boiling point (i.e., whether the transition is in the liquid regime of the fluid), and (b) whether or not long time bubble stability is affected by changing $\Theta_l$ (e.g., by changes in the gas solubility in the liquid which is a central parameter for stability [15]). The fact that this transition happens in the alcohols demonstrates that energy input to the gas by the Rayleigh-Plesset dynamics is sufficiently high for $\Lambda > 1$ to occur. Because bubbles in water are much brighter than bubbles in alcohol, we conjecture that this common situation is also underdamped. One may experimentally test the conjecture by carefully examining the temperature dependence of the onset of SBSL in air bubbles in water: At a critical driving pressure $P_a^*$ (depending on $\Theta_l$, $c_\infty$, etc.), the light turns on. By keeping the driving pressure fixed slightly above $P_a^*$ and then varying the liquid temperature $\Theta_l$, the light should turn off abruptly, in a fashion similar to Fig. 2.

There are other contributions to $\Lambda$ besides those mentioned above. An important loss mechanism is acoustic radiation through the wall, which leads to a reduction of the stored energy by a factor $D_a \approx 1 - (\rho_s c_{\text{gas}})/(\rho_l c_l)$ for every wave bounce in the bubble. When the bubble is fully expanded, $D_a \approx 1$ since $\rho_s \ll \rho_l$ and no losses occur. However, significant transmission occurs when the bubble is maximally compressed, where the gas density is of order the liquid density. Therefore, the light intensity should increase with the density of the liquid, and decrease with the density of the gas (with all other parameters held constant). Both these effects have been observed: experiments in alcohols [9] show a systematic increase in the light intensity with increasing density of the liquid. At room temperature the heaviest alcohol, pentanol, emits about 100 times more light than the lightest alcohol, ethanol. As for the gas in the bubble, the spectrum of xenon bubbles has a peak in the visible, whereas the lighter helium and argon bubbles show no such peak [5], suggesting a larger total intensity for the lighter gases.

Another experiment naturally interpreted is Weninger et al.'s observation [10] that the light emission from the bubble can be dipolar, not spherically symmetric. The dipole moment of the light emission shows significant correlations for delay times as long as 100 bubble oscillation cycles, many orders of magnitude longer than the pulse width. The present theory predicts that the gas density should have a significant dipole component, since the most easily excited mode shows this symmetry. This nonuniformity in the density causes differences in the index of refraction, producing an anisotropic light emission. The strength of the anisotropy depends on the size of the nonspherical component when the parametric instability saturates.

A basic difference between the present theory and the shock theory is that whereas the shock theory provides that successive flashes of light are completely independent, the present theory predicts strong correlations. Within our interpretation, the long lifetime of the dipole state measured by Weninger et al. [10] reflects these long time correlations.

To conclude, we have presented a mechanism for enhancement of shock focusing, and argued that this enhancement occurs in SBSL. The mechanism requires long time bubble stability, and thus does not work for transient bubble collapse in MBSL. By varying dissipation, the enhancement mechanism can be turned off, thus leading to abrupt jumps in light intensity [9]. The nonspherical gas motion implies a natural explanation for a long lived dipolar component of the light emission [10].

We thank S. Grossmann and L. Levitov for helpful discussions, and L. P. Kadanoff for comments on the manuscript. M. B. acknowledges an NSF postdoctoral fellowship; R. R. R. was partially supported by the NSF DMS Grant No. 9311438; M. B. and R. R. R. acknowledge the Sloan Fund of the School of Science at MIT. S. H. and D. L. acknowledge support by the DFG through its SFB185.